# On Concept of Creative Petri Nets

**Alexander Yu. Chunikhin**
Palladin Institute of Biochemistry
National Academy of Sciences of Ukraine
alexchunikhin61@gmail.com

**Abstract**. A new formalism of Petri nets, based on the adoption of the "position-arc-transition" triad and "transition-arc-position" triad as structure-forming units is introduced. In accordance with the Fusion principle, an analytical representation of Petri nets is developed. We propose the concept of Creative Petri Nets, which allows to implement structural changes in Petri net by procreation/deletion of structural units or complexes.

## 1. Introduction and Motivation

Petri nets (PN) are powerful modern modeling techniques for dynamic systems' simulation [2, 7, 8]. Parallelism and conflict accounting, discreteness and continuity, the ability to take into account inhibitory and associative (catalytic) influences on processes, fuzziness and stochasticity, functional equipment of positions, arcs and transitions - all these features make the Petri nets a flexible and effective tool for modeling a wide class of systems.

At the same time, the absence of the unified compact analytical representation that combines the structure, parameters and current state of an arbitrary Petri net makes it difficult to analyze its properties and functioning outside the graphical representation, especially with structural changes in the net. Besides, there is no general theory formalizing structural changes in the Petri net while its functioning, i.e. the aspect of network self-organization. The Open [1], reconfigurable [6], glued [3] and nested [4] Petri nets cover only certain aspects of the PN structural transformations, caused mainly by external control.

The development of a compact analytical representation of Petri nets, both elementary and functional, is the first goal of this work. The second goal is to create the foundations of the theory of PN self-organization, as a reflection of the prototype systems' self-organization, which foresees structural changes in the Petri net due to the occurrence of particular states in the process of its performance.

Both goals are associated with an attempt to create a uniform representation of the Petri net as a dynamic system with a variable structure.

## 2. Petri Nets Structural Units

### 2.1. Semantics

Components of the Petri net are traditionally considered as positions, transitions and arcs [7]. However, neither the components themselves (O, →, |), nor any of their permissible paired formations (O→, →|, |→, →O) have a complete sense in terms of the performing

elementary operations in the Petri net. Rather, positions, transitions and arcs can be called *proto-elements* of the Petri net.

**Proposition 1.** The *structural units* of the Petri net are defined as the "position-arc-transition" triad: O→| and the "transition-arc-position" triad: |→O.

"Position-arc-transition" - p-unit (O→|) – can be interpreted as a semantic unit-bundle object → action, resource → transformation. "Transition-arc-position" - t-unit (|→O) - can be interpreted as a semantic unit-bundle action → subject, process → resource.
They are the basic comprehensive structural formations, "bricks", on the basis of which the structure of an arbitrary Petri net is constructed and varied explicitly (without hidden assumptions).

## 2.2. Fusion Principle

Obviously, operating with triads instead of single proto-elements when building a Petri net requires different rules for composing the substructures into a structure.

**Proposition 2.** *Fusion Principle*. Any elementary Petri net can be designed from structural units (p-units and t-units) using fusion operators of two types: *p-fusion* is n-ary operator of composition of structural units by fusing their positions pf(_, ..., _); *t-fusion* is an n-ary operator of composition of structural units by fusing their transitions tf(_, ..., _).

For Petri nets providing for the possibility of multiple arcs it is reasonable to introduce another *n*-ary "mixed" *whole-fusion* operator: wf(_, ..., _), that is the fusion operator of both positions and transitions of the same structural unit. Such *n*-fold self-fusing leads to a corresponding increase in the multiplicity of arcs *n* times while maintaining the original position and transition.
Let us consider the principle of designing an elementary Petri net from structural units.
We will denote "C" as p-unit (O→|) and "I" as t-unit (|→O).
The coupling of the homogeneous structural units by fusion of their positions pf (_, _):

pf (C, C) = $_p$(C, C):      O→| ⇒ |←O→|,
                               * O→|

pf (I, I) = (I, I)$_p$.     |→O ⇒ |→O←|.
                             |→O *

The coupling of the homogeneous structural units by fusion of their transitions tf (_, _):

tf (C, C) = (C, C)$_t$:     O→| ⇒ O→|←O,
                            O→| *

tf (I, I) = $_t$(I, I) :     |→O ⇒ O←|→O.
                         * |→O

The coupling of the heterogeneous structural units:

tf (C, I) = CI: O→| * |→O  ⇒  O→|→O,

pf (I, C) = IC: |→O * O→|  ⇒  |→O→|.

The serial connection of heterogeneous units: O→|→O→|→O can be represented as

$$tf(pf(tf(C, I), C), I) = (((C, I)_t C)_p I)_t = CICI.$$

Expression 3C is interpreted as three separate p-units: O→|; O→|; O→|, and expression $C^2$ = wf(C, C) as a p-unit with a double arc:

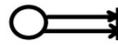

## 2.3. Functional Petri Nets

For functional Petri nets [5], we introduce the index form of the notation: position and transition numbers by lower indices; marking of positions (*m*), speeds of transitions (*v*) by top indices; thresholds of arcs (*k*) and transition delays (*d*) (if necessary) by brace parametric representation. In this case, it is advisable to introduce end-to-end numbering of positions and transitions for a more compact PN representation.
For example,

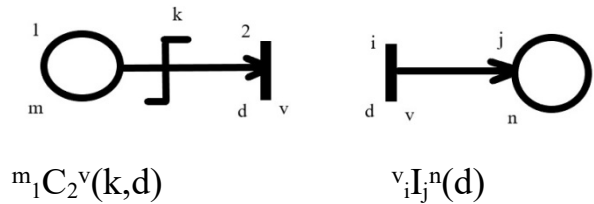

$$^m_1 C_2^v(k,d) \qquad ^v_i I_j^n(d)$$

We define the following rules for expressing a Petri net by formulas.
**R1**. Structural units are combined in the formula only by matching post-pre-indices.
**R2**. The general index for structural units can be bracketed: the pre-index is before the bracket, the post-index is after the bracket.

In PN-expression formula, the space remaining outside the index bracket will be denoted by a dot, for example, $_a C_k \circ\, _b C_k = (_a C., _b C.)_k$, where ∘ is the sign of both the units' composition before the fusion and fragments of the Petri net that cannot be fused in this net but included in it.
Besides, for functional Petri nets with inhibitory and associative arcs [5] we introduce additional notation for the corresponding p-units: "B" for unit with inhibitory arc and "A" for unit with associative arc. All the above presentation rules are fully true for them.

For example,

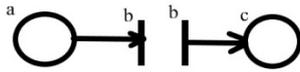

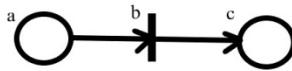

$$_aC_b \circ {_bI_c} \Rightarrow {_aC_bI_c}$$

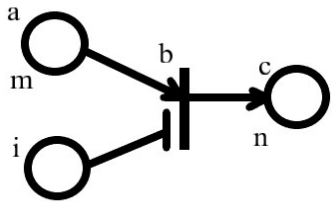

$$_aC_b \circ {_iB_b} \circ {_bI_c} \Rightarrow ({_aC_.}, {_iB_.})_bI_c$$

### 2.4. Defusion Principle

For the further presentation we need a contrary principle to the Fusion principle.

**Proposition 3.** *Defusion Principle*. Any Petri net can be subjected to complete or partial decomposition into its structural units by applying the *defusion operator*: Df (*expression*) to the entire PN or to any of its fragment.

For example,
$$PN = {_aI_bC_cI_e}; \quad Df(PN) = {_aI_b} \circ {_bC_c} \circ {_cI_e}; \quad Df(PN|c) = {_aI_bC_c} \circ {_cI_e}.$$

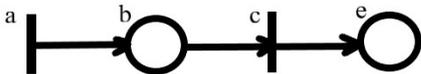

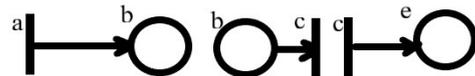

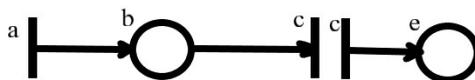

Since the use of the defusion operator is a preparatory step for structural transformations in the PN, the net is not performed in the defusion step.

### 3. Procreation/Deletion Principle

Most models based on Petri nets allow us to study the behavior of systems mainly by analyzing a change in the marking of the net, i.e. redistribution of the resource. However, in practice, there may be problems, processes, algorithms in which not only parametric, but also structural changes occur during the performance. Consequently, the PN-model that is adequate to such tasks should provide the ability to change its structure.

**Proposition 4**. A certain state resulting from the performance of a Petri net at some step is a prerequisite both the procreation of new structural units (their compositions) and the deletion of existing structural units of the Petri net. We will call this state *creative*.

Examples of creative states:
- the number of tokens in position $x$ is greater (less, equal) than a certain given value;
- the weighted sum of tokens in certain positions is greater (less, equal) than a certain given value;
For functional Petri nets:
- the speed (delay) of the transition $y$ is greater (less, equal) than a certain given value.

**Proposition 5**. *Procreation/Deletion Principle*. The elementary procreation (deletion) in a PN is defined as the fact of the introduction (removal) of a new p-unit or t-unit into its structure due to the creative state of this net.

Presentation format: (condition) <*expression*> - for procreation; (condition) >*expression*< - for deletion.
For example, let a Petri net $PN = (_aI_bC., _1A.)_cI_e$ be given

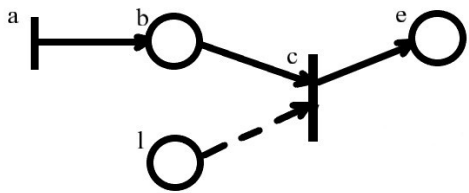

Example of procreation: $(\text{if}\_) < {}_cI_k> \Rightarrow PN' = (_aI_bC., {}_1A.)_c(.I_e, .I_k)$

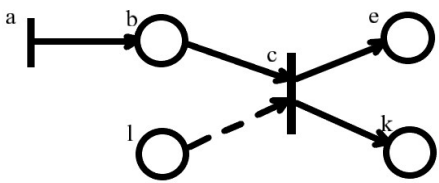

Example of deletion: $(\text{if}\_) >_aI_b< \Rightarrow PN'' = (_bC., {}_1A.)_c(.I_e, .I_k)\|$

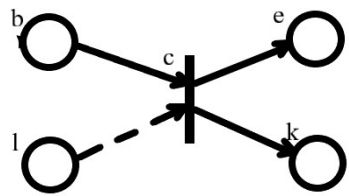

The advantage of the proposed approach is the absence of the need for an additional indication of the "address" where the procreated unit is implemented or removed from. This is automatically determined by the lower unit indices. Changes in the PN formula are made in concordance with the principle of fusion (deletion) according to the R1, R2 rules. Not only individual structural units but also their associated aggregates - *complexes* (PN fragments) can be procreated (deleted). We will also assume that the structural unit is a singular complex.

**Proposition 6**. The fact of the introduction (withdrawal) of a PN-complex or several complexes into its structure due to the creative state of this net will be called the *complex procreation (deletion)* in the PN.

For example, $PN = (_aI_bC., _lA.)_cI_e$

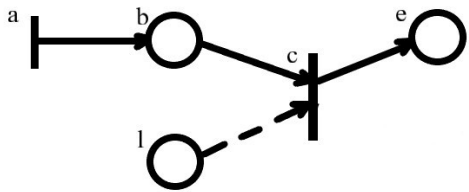

$(if\_) <_cI_kC_f(.I_e, .I_g)> \Rightarrow PN' = (_aI_bC., _lA.)_c(.I_e, .I_kC_f(.I_l, .I_g))$

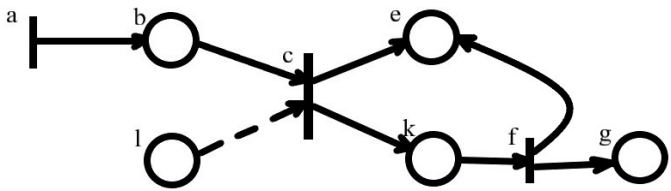

Obviously, the processes of procreation (deletion) can not coincide with the performance for net elements, that is, at the time of procreation (deletion) the corresponding sections of PN are not performed.
Possible options:
- "greedy" - the areas of procreation (deletion) and all transitions that are allowed to be completed but are not involved in the creation are performed;
- "sequential" - at the step where the procreations (deletions) are present, only the procreations (deletions) with the corresponding change in the net structure and its marking are performed. In the next step all allowed transitions of the updated network are performed.

## 4. Marking

We define two types of procreation - resource-free and resource-intensive. Upon resource-free procreation, the current marking and accordingly, the total PN resource does not change. Upon resource-intensive - a certain amount of resource (tokens) from specific positions is spent on creating each PN unit. Part of the resource can be used for structural creation and another part can be used to fill positions in the created units.

Example of resource-intensive procreation:
$$PN = {}^5_aC_bI_c^2,$$

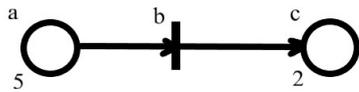

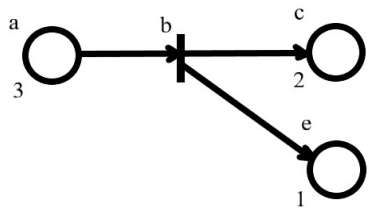

$$PN' = {}^3_aC_b(.I_c^2, .I_e^1)$$

Accordingly, when deleting structural units of a network, the following options are possible:
1) the deletion does not change the current marking of the remaining part of the network;
2) the deletion of each structural unit requires a certain amount of resource, which is withdrawn from one or several positions in a given ratio;
3) the deletion of each structural unit "releases" a certain amount of the resource, placed then in certain positions. At the same time, the choice of positions-"receivers" can be made both according to the rule of the nearest neighbor (predecessor or successor), or by another rule.

For example, $PN = {}^5_aC_b(.I_c^4, .I_e^3)$, $(if\_) >_bI_e^3<| \; m_c' = m_c + m_e$:

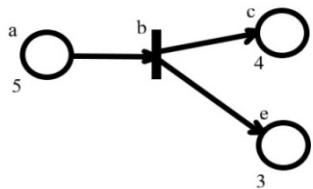

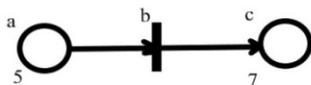

$$PN' = {}^5_aC_bI_c^7$$

A separate consideration deserves the case when the resource released upon deletion is used for procreation at the same step.

For example, $PN = {}^5_aC_b(.I_c^4, .I_e^3)$, $(if\_) >_bI_e^3<| <_cC_f>, \; m_c' = m_c + 2_e$:

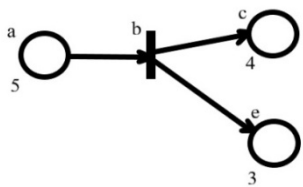

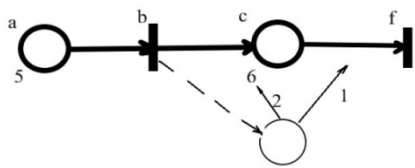

$$PN' = {}^5_aC_bI_c^6C_f$$

It is also necessary to point out certain changes in the semantics of the PN after procreation (deletion).

The procreation of a p-unit or the deletion of a t-unit at the input of a net changes its entry semantics from "process" (producing) to resource. And vice versa when a t-unit is procreated or a p-unit is deleted at the net input. Accordingly, upon procreation (deletion) of the final units, the semantics of the end of the net also changes to the opposite: the stock interpretation is replaced by the accumulative, the accumulative one - by the stock one. Changes in the semantics of the entry/end during the procreation (deletion) of the complex will be determined by the type of the first (last) unit in the complex.

## 5. Conclusion

Based on the introduction of the Petri net structural units' concept and the Fusion principle, a new method of formalization has been developed. It allows to compactify the representation of both elementary and functional Petri nets.

We proposed the concept of Creative Petri Nets in which it becomes possible both to procreate new structural units (complexes) of the Petri net and implement them in a given net as well as to delete existing structural units (complexes) in the net.

Two types of creativity are defined – resource-free and resource-intensive. Resource-free does not use the network resource to procreate / delete structural units. Resource-intensive requires the cost of a certain part of an available net resource for procreation and possibly deletion. It is possible to redistribute the resource released upon deletion (both structural units and tokens) over the remaining network positions.

The proposed concept enables to use the Petri nets for modeling complex dynamic systems with a variable structure.

It is advisable to study the applicability of this approach to formalize procreation/deletion in nested Petri nets [4].

## Acknowledgements

The author wants to thank Marina Sviatnenko for the useful discussions.